\documentclass[twocolumn]{aastex701}

\newcommand\target{EBG-1}
\newcommand{\fesc}{$f_{\rm esc}$}

\newcommand{\hb}{\hbox{H$\beta$}}

\newcommand{\oiii}{\hbox{O\,{\sc iii}}}
\newcommand{\oii}{\hbox{O\,{\sc ii}}}

\newcommand{\hii}{\hbox{H\,{\sc ii}}}

\newcommand{\ciii}{\hbox{C\,{\sc iii}}}

\usepackage{CJK}

\begin{document}
\begin{CJK*}{UTF8}{gbsn}

\title{Can Blue UV-Bright Galaxies Drive Reionization at Cosmic Dawn? An \(M_{\rm UV}=-19.6\) Galaxy with \(f_{\rm esc}\approx0.85\) at $z=9.244$}

\author[0000-0003-0111-8249]{Yunjing Wu}
\altaffiliation{Kavli IPMU Special Postdoctoral Fellow}
\affiliation{Kavli Institute for the Physics and Mathematics of the Universe (WPI), The University of Tokyo Institutes for Advanced Study, The University of Tokyo, Kashiwa, Chiba 277-8583, Japan}
\affiliation{Center for Data-Driven Discovery, Kavli IPMU (WPI), UTIAS, 
The University of Tokyo, Kashiwa, Chiba 277-8583, Japan}
\email[show]{yunjing.wu@ipmu.jp}

\author[0009-0003-4133-0292]{Sijia Cai}
\affiliation{Department of Astronomy, Tsinghua University, Beijing 100084, China}
\email{caisj23@mails.tsinghua.edu.cn}

\author[0000-0003-3307-7525]{Yongda Zhu}
\thanks{JASPER Scholar}
\affiliation{Steward Observatory, University of Arizona, 933 North Cherry Avenue, Tucson, AZ 85721, USA}
\email{yongdaz@arizona.edu}

\author[0000-0002-1049-6658]{Masami Ouchi}
\affiliation{National Astronomical Observatory of Japan, 2-21-1 Osawa, Mitaka, Tokyo 181-8588, Japan}
\affiliation{Institute for Cosmic Ray Research, The University of Tokyo, 5-1-5 Kashiwanoha, Kashiwa, Chiba 277-8582, Japan}
\affiliation{Department of Astronomical Science, SOKENDAI (The Graduate University for Advanced Studies), 2-21-1 Osawa, Mitaka, Tokyo, 181-8588, Japan}
\affiliation{Kavli Institute for the Physics and Mathematics of the Universe (WPI), The University of Tokyo Institutes for Advanced Study, The University of Tokyo, Kashiwa, Chiba 277-8583, Japan}
\email{ouchims@icrr.u-tokyo.ac.jp}

\author[0000-0002-6047-430X]{Yuichi Harikane}
\affiliation{Institute for Cosmic Ray Research, The University of Tokyo, 5-1-5 Kashiwanoha, Kashiwa, Chiba 277-8582, Japan}
\email{hari@icrr.u-tokyo.ac.jp}

\author[0000-0001-8467-6478]{Zheng Cai}
\affiliation{Department of Astronomy, Tsinghua University, Beijing 100084, China}
\email{zcai@mail.tsinghua.edu.cn}

\author[0000-0001-5940-338X]{Mengtao Tang}
\affiliation{Tsung-Dao Lee Institute, Shanghai Jiao Tong University, 1 Lisuo Road, Shanghai 201210, People’s Republic of China}
\affiliation{School of Physics and Astronomy, Shanghai Jiao Tong University, 800 Dongchuan Road, Shanghai 200240, People’s Republic of China}
\email{mengtao.tang@sjtu.edu.cn}

\author[0000-0001-9299-5719]{Khee-Gan Lee}
\affiliation{Kavli Institute for the Physics and Mathematics of the Universe (WPI), The University of Tokyo Institutes for Advanced Study, The University of Tokyo, Kashiwa, Chiba 277-8583, Japan}
\affiliation{Center for Data-Driven Discovery, Kavli IPMU (WPI), UTIAS, The University of Tokyo, Kashiwa, Chiba 277-8583, Japan}
\email{kglee@ipmu.jp}

\author[0000-0002-0000-6977]{John D. Silverman}
\affiliation{Kavli Institute for the Physics and Mathematics of the Universe (WPI), The University of Tokyo Institutes for Advanced Study, The University of Tokyo, Kashiwa, Chiba 277-8583, Japan}
\affiliation{Center for Data-Driven Discovery, Kavli IPMU (WPI), UTIAS, The University of Tokyo, Kashiwa, Chiba 277-8583, Japan}
\affiliation{Department of Astronomy, Graduate School of Science, The University of Tokyo, 7-3-1 Hongo, Bunkyo, Tokyo 113-0033, Japan}
\affiliation{Center for Astrophysical Sciences, Department of Physics \& Astronomy, Johns Hopkins University, Baltimore, MD 21218, USA}
\email{silverman@ipmu.jp}

\author[0000-0002-4622-6617]{Fengwu Sun}
\affiliation{Department of Astronomy, Westlake University, Hangzhou 310030, Zhejiang Province, China}
\affiliation{Center for Astrophysics $|$ Harvard \& Smithsonian, 60 Garden St., Cambridge, MA 02138, USA}
\email{sunfengwu@westlake.edu.cn}

\author[0000-0003-3310-0131]{Xiaohui Fan}
\affiliation{Steward Observatory, University of Arizona, 933 North Cherry Avenue, Tucson, AZ 85721, USA}
\email{xfan@arizona.edu}

\author[0000-0002-2178-5471]{Zuyi Chen}
\affiliation{Cosmic Dawn Center (DAWN)}
\affiliation{Niels Bohr Institute, University of Copenhagen, Jagtvej 128, 2200 Copenhagen N, Denmark}
\email{zuyi.chen@nbi.ku.dk}

\author[0009-0000-1999-5472]{Minami Nakane}
\affiliation{Institute for Cosmic Ray Research, The University of Tokyo, 5-1-5 Kashiwanoha, Kashiwa, Chiba 277-8582, Japan}
\affiliation{Department of Physics, Graduate School of Science, The University of Tokyo, 7-3-1 Hongo, Bunkyo, Tokyo 113-0033, Japan}
\email{nakanem@icrr.u-tokyo.ac.jp}  

\author[orcid=0009-0006-6763-4245]{Hiroto Yanagisawa}
\affiliation{Institute for Cosmic Ray Research, The University of Tokyo, 5-1-5 Kashiwanoha, Kashiwa, Chiba 277-8582, Japan}
\affiliation{Department of Physics, Graduate School of Science, The University of Tokyo, 7-3-1 Hongo, Bunkyo, Tokyo 113-0033, Japan}
\email{yana@icrr.u-tokyo.ac.jp}

\author[0000-0002-8192-8091]{Angela Adamo}
\affiliation{Department of Astronomy, The Oskar Klein Centre, Stockholm University, AlbaNova, SE-10691 Stockholm, Sweden}
\email{antonella.nota@issibern.ch}

\author[0000-0002-1620-0897]{Fuyan Bian}
\affiliation{European Southern Observatory, Alonso de C\'ordova 3107, Casilla 19001, Vitacura, Santiago 19, Chile}
\affiliation{Chinese Academy of Sciences South America Center for Astronomy, National Astronomical Observatories, CAS, Beijing 100101, China}
\email{fuyan.bian@eso.org}

\author[0000-0002-6184-9097]{Jaclyn B. Champagne}
\affiliation{Space Telescope Science Institute, 3700 San Martin Drive, Baltimore, MD 21218, USA}
\email{jchampagne@stsci.edu}

\author[0000-0001-7410-7669]{Dan Coe}
\affiliation{Space Telescope Science Institute, 3700 San Martin Drive, Baltimore, MD 21218, USA}
\affiliation{Association of Universities for Research in Astronomy (AURA), Inc.~for the European Space Agency (ESA)}
\affiliation{Department of Physics and Astronomy, The Johns Hopkins University, 3400 N Charles St. Baltimore, MD 21218, USA}
\email{dcoe@stsci.edu}

\author[0000-0001-7201-5066]{Seiji Fujimoto}
\affiliation{Department of Astronomy, The University of Texas at Austin, Austin, TX 78712, USA}
\email{seiji.fujimoto@utoronto.ca}

\author[0000-0002-6021-7020]{Enrico Garaldi}
\affiliation{Kavli Institute for the Physics and Mathematics of the Universe (WPI), The University of Tokyo Institutes for Advanced Study, The University of Tokyo, Kashiwa, Chiba 277-8583, Japan}
\affiliation{Center for Data-Driven Discovery, Kavli IPMU (WPI), UTIAS, The University of Tokyo, Kashiwa, Chiba 277-8583, Japan}
\affiliation{Institute for Fundamental Physics of the Universe, via Beirut 2, 34151 Trieste, Italy }
\email{egaraldi@ipmu.jp}

\author[0000-0003-4512-8705]{Tiger Yu-Yang Hsiao} 
\affiliation{Department of Astronomy, The University of Texas at Austin, 2515 Speedway, Austin, Texas 78712, USA}
\affiliation{Cosmic Frontier Center, The University of Texas at Austin, Austin, TX 78712}
\email{tiger.hsiao@utexas.edu}

\author[orcid=0000-0002-5768-738X,sname='Jin']{Xiangyu Jin}
\affiliation{Department of Astronomy, University of Michigan, 1085 S. University Ave., Ann Arbor, MI 48109, USA}
\email{jxiangyu@umich.edu}  

\author[0000-0001-6251-649X]{Mingyu Li}
\affiliation{Department of Astronomy, Tsinghua University, Beijing 100084, China}
\affiliation{Kavli Institute for Cosmology, University of Cambridge, Madingley Road, Cambridge CB3 0HA, UK}
\affiliation{Cavendish Laboratory, University of Cambridge, 19 JJ Thomson Avenue, Cambridge CB3 0HE, UK}
\email{lmytime@hotmail.com}

\author[0000-0002-2725-302X]{Yongming Liang} 
\affiliation{Institute for Cosmic Ray Research, The University of Tokyo, 5-1-5 Kashiwanoha, Kashiwa, Chiba 277-8582, Japan} 
\affiliation{National Astronomical Observatory of Japan, 2-21-1 Osawa, Mitaka, Tokyo 181-8588, Japan}
\email{ymliang@icrr.u-tokyo.ac.jp}

\author[0000-0001-6052-4234]{Xiaojing Lin}
\affiliation{Department of Astronomy, Tsinghua University, Beijing 100084, China}
\email{xiaojinglin.astro@gmail.com}

\author[0000-0002-3464-4424]{Guan-Fu Liu (刘观福)}
\affiliation{Department of Astronomy, Tsinghua University, Beijing 100084, China}
\email{lgf22@mails.tsinghua.edu.cn}  

\author[0000-0001-7557-9713]{Junjie Mao}
\affiliation{Department of Astronomy, Tsinghua University, Beijing 100084, China}
\email{jmao@mail.tsinghua.edu.cn}

\author[0000-0003-3484-399X]{Masamune Oguri}
\affiliation{Center for Frontier Science, Chiba University, 1-33 Yayoicho, Inage, Chiba 263-8522, Japan}
\affiliation{Department of Physics, Graduate School of Science, Chiba University, 1-33 Yayoicho, Inage, Chiba 263-8522, Japan}
\email{masamune.oguri@chiba-u.jp}

\author[0000-0002-1499-6377]{Takafumi Tsukui}
\affiliation{Kavli Institute for the Physics and Mathematics of the Universe (WPI), The University of Tokyo Institutes for Advanced Study, The University of Tokyo, Kashiwa, Chiba 277-8583, Japan}
\affiliation{Center for Data-Driven Discovery, Kavli IPMU (WPI), UTIAS, 
The University of Tokyo, Kashiwa, Chiba 277-8583, Japan}
\email{tsukuitk23@gmail.com}

\author[0000-0002-7633-431X]{Feige Wang}
\affiliation{Department of Astronomy, University of Michigan, 1085 S. University Ave., Ann Arbor, MI 48109, USA}
\email{fgwang@umich.edu}

\author[0000-0001-5287-4242]{Jinyi Yang}
\affiliation{Department of Astronomy, University of Michigan, 1085 S. University Ave., Ann Arbor, MI 48109, USA}
\email{jyyangas@umich.edu}

\begin{abstract}
We revisit \target, a gravitationally lensed galaxy at $z = 9.244$ using newly obtained deep (14\,hr) JWST/NIRSpec G235M ($R \sim 1000$) spectroscopy. 
\target\ is one of the bluest galaxies observed at $z > 9$.
Along with its remarkably blue continuum of $\beta = -2.99 \pm 0.15$, 
We find very weak nebular emission of \target.
We obtain a $3\sigma$ upper limit of EW $< 8.4$\,\AA\ for \ciii], a marginal H$\beta$ measurement (EW $= 256.2 \pm 122.2$\,\AA), and a $5\sigma$ detection of [\oiii]\,$\lambda 5007$ (EW $= 882.8 \pm 180.6$\,\AA).
These lines suggest the element abundance of $12 + \log({\rm O/H}) = 7.16 \pm 0.25$ ($\simeq 3\%\,Z_{\odot}$) and $\log({\rm C/O}) < -0.88$ ($3\sigma$), placing \target\ on the C/O--metallicity sequence traced by lower-redshift galaxies.
To constrain the escape fraction (\fesc), we compare the observed UV slope and emission-line equivalent widths with synthetic spectra generated using {\tt BEAGLE}.
The models favor $f_{\rm esc}\gtrsim0.75$, with $f_{\rm esc}\approx0.85$ providing the best overall agreement with the observations.
These results suggest that relatively UV-bright galaxies like \target\ can achieve the exceptionally high ionizing efficiencies supported by late reionization scenarios.
\end{abstract}


\section{Introduction}
\label{sec:01_intro}

The completion of cosmic reionization of hydrogen by $z \sim 5.5$ marks the last major phase transition of the Universe \citep[e.g.,][]{Ouchi2020, Fan2023,Stark2026}.
The recent discovery of abundant faint AGNs has renewed interest in their potential contribution to hydrogen reionization, given their role in producing the hard ionizing background at $z<4$ \citep{Worseck2019,Garaldi2019,Madau2024}.
However, star-forming galaxies remain the leading candidates for supplying the most of the ionizing photons \citep{Atek2024, Jiang2025}.

Which galaxies supplied most of the ionizing photons remains debated.
Many models identify dwarf galaxies as the primary contributors of Lyman-continuum (LyC) photons, due to their large abundance and potentially efficient escape of ionizing radiation~\citep{Yeh2023,Simmonds2024,Saxena2024}.
Others argue that more massive galaxies dominate, as their deeper gravitational potentials sustain efficient star formation and consequently greater LyC photon production \citep{Naidu2020,Harikane2023}.
Central to this debate is the escape fraction \fesc\ of LyC photons that leak from galaxies into the intergalactic medium (IGM).
Together with the ionizing photon production efficiency $\xi_{\rm ion}$ and the UV luminosity density $\rho_{\rm UV}$, it sets the ionizing emissivity of galaxies, $\dot{n}_{\rm ion} = $\fesc$\,\xi_{\rm ion}\,\rho_{\rm UV}$ \citep{Robertson2022}.
Late reionization scenarios place particularly strong demands on this photon budget and can require highly efficient LyC escape from the galaxies driving the transition \citep{Nakane2024,Tang2024,Umeda2024,Umeda2026,Kageura2025,Meyer2025,Mason2026,Napolitano2024,Napolitano2026}.
A population of efficient leakers could contribute disproportionately to this budget even if most galaxies have low \fesc.
Identifying such galaxies therefore tests whether the efficient ionizing sources invoked by these scenarios exist.

The high LyC opacity of the IGM during reionization prevents direct measurements of \fesc\ and necessitates indirect diagnostics.
The rest-frame ultraviolet (UV) continuum slope $\beta$ ($f_{\lambda} \propto \lambda^{\beta}$) provides one such probe.
A higher \fesc\ reduces the nebular continuum emission that reddens the spectra of young stellar populations and can therefore produce a bluer UV slope \citep{Chisholm2022, Papovich2023}.
This motivates searches for galaxies with extremely blue UV slopes ($\beta \lesssim -3.0$) at $z > 6$ as promising candidates for metal-poor and nearly dust-free systems with substantial LyC leakage \citep{Jiang2020,Wang2024, Messa2025, Jecmen2026}.
Combining their UV slopes with nebular emission-line measurements can test whether high \fesc\ is required to explain their spectral properties.

Recent JWST imaging has extended the search for such blue galaxies out to $z \gtrsim 9$ \citep{Cullen2024, Topping2024, Donnan2025, Cheng2026}.
Spectroscopic confirmation, however, remains limited to a small number of systems \citep{Witstok2025, Witstok2026}.
Among these, Extreme Blue Galaxy 1 (hereafter \target) is a luminous ($M_{\rm UV} = -19.6$) galaxy at $z = 9.244$ lensed by the MACS0647 cluster.
It stands out for having one of the bluest UV slopes yet measured at this redshift \citep{Yanagisawa2025}.
Existing prism spectra (R$\sim100$) confirm the redshift and the blue slope, but could not to constrain the nebular line strengths, metallicity, or ISM conditions that would establish whether such a galaxy is a genuine LyC leaker.
Deep, medium-resolution grating spectroscopy is therefore required to perform this test.

In this Letter, we present deep (14 hr) JWST/NIRSpec G235M spectroscopy of \target.
These data resolve the strong rest-frame optical emission lines and constrain the nebular conditions, gas-phase metallicity, and UV slope of the galaxy.
This Letter is organized as follows.
We describe the observations and data reduction in Section~\ref{sec:02_obs}.
In Section~\ref{sec:03_res}, we present the emission line measurements and the derived elemental abundances.
We then present the stellar population properties derived from SED fitting, and compare our emission-line diagnostics with the UV slope to infer the escape fraction and the ISM conditions in Section~\ref{sec:04_disc}.
Finally, we summarize our conclusions in Section~\ref{sec:05_sum}.
Throughout this Letter, we assume a flat $\Lambda$CDM cosmology with $\Omega_{\rm M} = 0.3$, $\Omega_{\Lambda} = 0.7$, and $H_0 = 70\ {\rm km\ s^{-1}\ Mpc^{-1}}$.

\begin{figure*}
    \centering
    \includegraphics[width=0.98\linewidth]{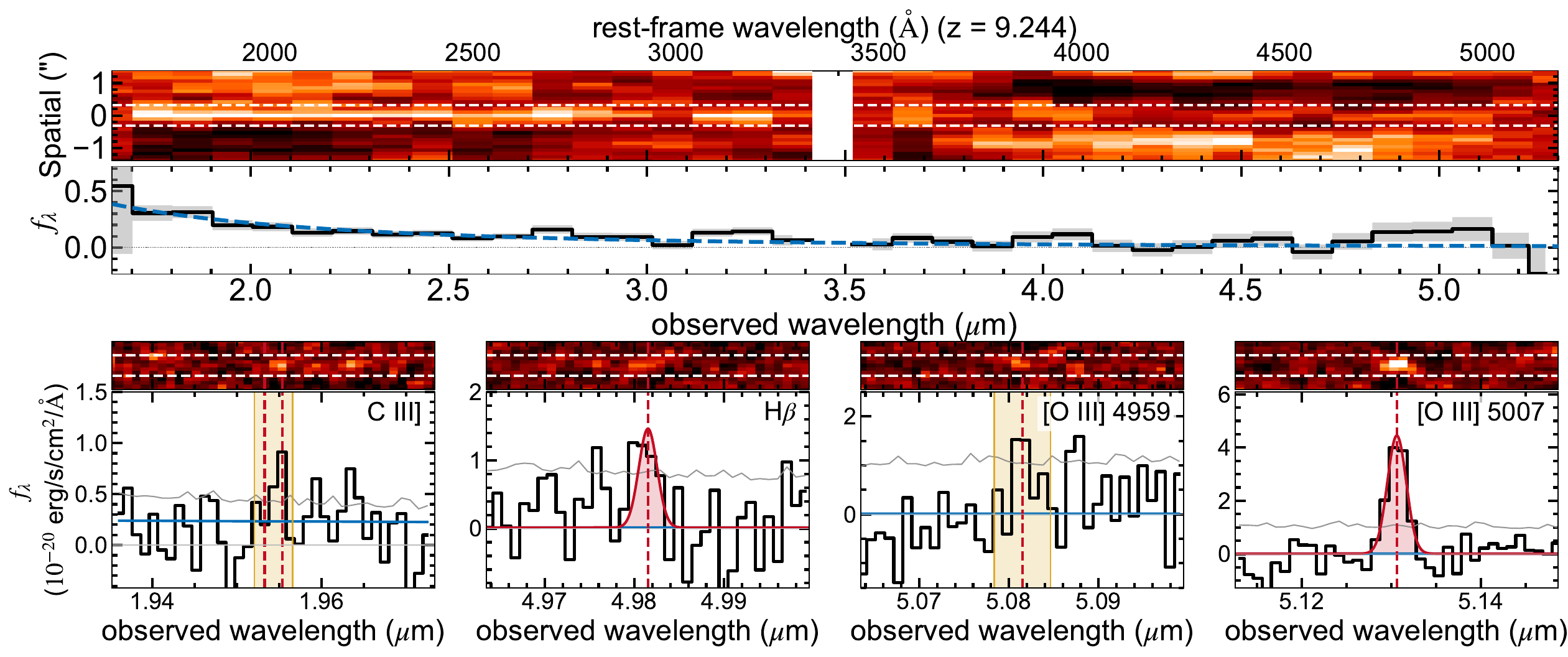}
    \caption{
    JWST/NIRSpec G235M spectrum of \target, with wavelength coverage extended to $\lambda_\mathrm{obs} \approx 5.5\,\mu$m. \textbf{Top:} The upper sub-panel shows the 2D spectrum (binned by a factor of 100) to highlight the continuum detection. The extraction aperture is indicated by the dashed white lines and the detector gap masked near $3.4\,\mu$m. The lower sub-panel shows the 1D extracted spectrum, with the black histogram giving the binned flux and the grey shaded region the $1\sigma$ uncertainty. The blue dashed curve is the best-fit power-law continuum with fixed UV slope $\beta = -2.99$. \textbf{Bottom (left to right):} Zoom-ins at the native $R \sim 1000$ resolution on the rest-frame UV and optical emission lines. Each panel shows the 2D spectra in the upper sub-panel, with the extraction aperture again marked by the dashed white lines. In the lower sub-panel, the black histogram is the 1D spectrum, the grey curve shows the $1\sigma$ noise level, while the blue horizontal line is the fitted continuum, and the red dashed vertical line marks the expected line centroid at the systemic redshift. For detected lines and tentative detections (SNR $> 2$), the red shaded curve shows the best-fit Gaussian profile. For non-detections, the yellow shaded band indicates the wavelength window used for flux integration and upper-limit estimation. The systemic redshift of \target, $z = 9.244$, is derived from [O\,\textsc{iii}]\,$\lambda 5007$ and is consistent with the value reported by \citet{Yanagisawa2025}.}
    \label{fig:EBG1_spec}
\end{figure*}

\section{Observations and Data Processing}
\label{sec:02_obs}
In this work, we present an observing program designed to characterize bright, spectroscopically confirmed high-redshift galaxies in the MACS0647 field.
The program combines imaging and spectroscopy from the Hubble Space Telescope (HST) and the James Webb Space Telescope (JWST).
Primary targets are selected from sources with existing spectroscopic redshifts.
Filler targets are drawn from deep HST and JWST imaging of the same field.

Here, we first focus on \target\ that was identified from a parent sample of 863 galaxies at $4 < z < 14$.
This parent sample was assembled from archival JWST/NIRSpec PRISM observations carried out under major JWST GTO, ERS, and GO programs.
The source was first identified photometrically as MACS0647-z9-20158 ($z_{\rm phot} = 9.5$) by \citet{McLeod2024} in GO 1433 \citep[PI: Dan Coe;][]{Coe2021jwst.prop.1433}.
It was later observed as a filler target in the MACS0647 field \citep{Worku2026}.
\citet{Yanagisawa2025} reported a spectroscopic redshift of $z = 9.25$ and a prism-derived, exceptionally blue UV slope of $\beta = -2.99 \pm 0.15$.

\subsection{JWST NIRCam imaging, photometry, and lens model}
The six-band NIRCam imaging data are obtained from the Cycle-1 program GO~1433 \citep[PI: Coe;][]{Coe2021jwst.prop.1433, Hsiao2023}.
We summarize the reduction here and refer the reader to \citet{Guo2026} for full details.
The reduction is based on the JWST pipeline \citep[v1.18.1;][]{Bushouse2022} with CRDS context {\tt jwst\_1364.pmap}, supplemented by customized steps for $1/f$ noise subtraction, sky background subtraction, and astrometric correction.
Astrometry is registered to the CLASH (Cluster Lensing And Supernova survey with Hubble) catalog \citep{Coe2013}.
The mosaics are then co-added to a pixel scale of 0\farcs03 in all bands.

We perform PSF-matched aperture photometry on the calibrated mosaics using model PSFs from {\tt STPSF} \citep{Perrin2014}.
We adopt a $d=0\farcs35$ aperture, following \citet{Yanagisawa2025} and match all bands to the F444W PSF.
Aperture corrections are derived from the same model PSF following \citet{Ji2024}.
Photometric uncertainties are estimated as the standard deviation of $10{,}000$ apertures randomly dropped across source-free regions of the mosaics.
The measured apparent magnitude of F200W is $m_{\rm F200W} = 27.5$\,mag.
We note that, although using the same aperture, our photometry is brighter by $\sim0.2$\,mag than that reported in \citet{Yanagisawa2025}, a difference we attribute to the PSF-matching, resampling for the longer wavelength images and aperture-correction procedures. The full photometry is summarized in Table~\ref{tab:phot} in Appendix~\ref{Appendix_phot}.

For the lensing magnification correction, we adopt the model of \citet{Nishida2025}, which is constructed with the {\tt glafic} software \citep{Oguri2010,Oguri2021} using 86 multiple images from 26 sources.
The inferred magnification of EBG-1 is $\mu = 2.75$ \citep[see also][]{Worku2026}.

\subsection{Deep NIRSpec G235M Spectroscopy}
We present new deep JWST/NIRSpec G235M/F170LP spectroscopy (R$\sim$1000) of \target\ from the Cycle-4 program GO~7390 \citep[PI: Y. Wu;][]{Wu2025}.
The observations  were carried out in March 2026.
The program targeted $>$100 unique sources in the MACS0647 field across three independent MSA configurations, making it one of the deep NIRSpec/G235M MOS surveys conducted to date.
Each configuration reached an integration time of 25{,}209.6\,s ($\approx 7$\,hr).
\target\ was covered by two of the three configurations, providing a total on-source exposure time of $14$\,hr.
The G235M spectra were reduced with the \texttt{msaexp}\footnote{\url{https://github.com/gbrammer/msaexp/blob/main/docs/examples/spectral-pipeline-2025.ipynb}} pipeline \citep[0.1.dev1+g44392ac,][]{Brammer2023zndo...8319596B} to recover the extended wavelength coverage.
We processed the raw NIRSpec data through Stage~1 of the standard STScI \texttt{jwst} pipeline \citep[v1.17.1,][]{Bushouse2025zndo..14597407B}, adopting the CRDS context \texttt{jwst\_1481.pmap}, to generate the \texttt{rate}-level products. Stage~2 and subsequent processing were performed with \texttt{msaexp}. Prior to Stage~2 calibration, we applied a $1/f$-noise correction and rescaled the \texttt{VAR\_RNOISE} arrays using source-free detector regions. The spectra were then calibrated with the extended wavelength-dependent reference files, background-subtracted using nod differences, drizzled, and optimally extracted.

\section{Results}
\label{sec:03_res}
In this section, we characterize the emission lines and the diagnostics of \target.
We summarized the emission-line measurements in Sec.~\ref{sec:03_1_EMLINE} and the Elemental abundanced derived from emission lines in Sec.~\ref{sec:03_2_Elemental_abundance}.

\subsection{Emission-Line and Continuum Measurements}
\label{sec:03_1_EMLINE}
Line fluxes and rest-frame equivalent widths (EWs) are measured from the extracted 1D spectra with the following method. 
We first estimate the underlying continuum by binning the spectrum by 100 pixels and fitting a power law $f_\lambda \propto \lambda^{\beta}$ with a fixed slope $\beta = -2.99$ reported by \citet{Yanagisawa2025} to the binned points at $\lambda < 3.3\,\mu$m; the best-fit power law is then extrapolated to the wavelengths of all lines.
Emission lines are measured by fitting single Gaussians to the continuum-subtracted spectrum. 
The systemic redshift is estimated by the centroid of \oiii]\,$\lambda5007$ as $z_{\rm sys} = 9.2443 \pm 0.0005$, which is consistent with $z_{\rm spec} = 9.25$ reported under the prism (R100) reported by \citet{Yanagisawa2025}.
We fit \oiii]\,$\lambda5007$ with a free Gaussian to establish a reliable intrinsic line width convolved with the line-spread function, obtaining $\sigma_v \approx 62\,\mathrm{km\,s^{-1}}$ (FWHM\,$\approx 145\,\mathrm{km\,s^{-1}}$). 
This width keeps fixed for all lines, with only the line amplitudes left as free parameters.
The non-detection of [\oiii]$\lambda4959$ is consistent with the theoretical [\oiii]$\lambda5007/\lambda4959$ flux ratio of $\approx3$.
We modeled \ciii] doublets as a sum of two Gaussians sharing this common width. 
We estimated the line flux uncertainties based Monte Carlo resampling: the spectrum is perturbed 1000 times by the $1\sigma$ error array following \citet{Tang2026}.
The signal-to-noise ratio (S/N) is defined as the line flux divided by this
uncertainty.
We report $3\sigma$ upper limits for non-detections with $\mathrm{S/N} < 2$.
Equivalent widths are computed as $\mathrm{EW_{obs}} = F_\mathrm{line}\,/\,f_{\lambda,\mathrm{cont}}$, where $f_{\lambda,\mathrm{cont}}$ is the continuum flux density evaluated at the line center, with the rest-frame EW as $\mathrm{EW_{rest}} = \mathrm{EW_{obs}}\,/\,(1+z)$.
Table~\ref{tab:ebg1_lines} summarizes the resulting measurements.

\begin{deluxetable}{ccc}
\tablecaption{\label{tab:ebg1_lines}Emission lines, line ratios, and physical properties estimated for \target.}
\tablewidth{\columnwidth}
\tablehead{
\colhead{Line} & \colhead{Flux} & \colhead{EW} \\
\colhead{}     & \colhead{[$10^{-20}\,$erg/s/cm$^2$]} & \colhead{[\AA]}
}
\startdata
C\,{\sc iii}]\,$\lambda\lambda$1907,1909  & $<19.98$ (3$\sigma$)   & $<8.4$ \\
H$\beta$                                  & $37.18\pm17.73$        & $256.2\pm122.2$ \\
{}[O\,{\sc iii}]\,$\lambda$4960           & $<63.46$ (3$\sigma$)   & $<464.2$ \\
{}[O\,{\sc iii}]\,$\lambda$5007           & $117.27\pm23.99$       & $882.8\pm180.6$ \\ \hline
\multicolumn{3}{c}{Emission Line Ratios} \\ \hline
R3 = [O\,{\sc iii}]\,$\lambda$5007\ $/$ H$\beta$                              & \multicolumn{2}{c}{$3.15\pm1.64$} \\
C\,{\sc iii}]\,$\lambda\lambda$1907,1909\ $/$ [O\,{\sc iii}]\,$\lambda$5007   & \multicolumn{2}{c}{$<0.17$ (3$\sigma$)} \\ \hline
\multicolumn{3}{c}{Physical Properties} \\ \hline
12+log(O/H)            & \multicolumn{2}{c}{$7.16 \pm 0.25$} \\
${\rm C^{2+}/O^{2+}}$  & \multicolumn{2}{c}{$<0.13$} \\
log(C/O)               & \multicolumn{2}{c}{$<-0.88$} \\
\enddata
\tablecomments{Line fluxes measured from the NIRSpec G235M spectrum and the extended wavelength. Upper limits are 3$\sigma$.}
\end{deluxetable}

\subsection{Elemental Abundance Measurements}
\label{sec:03_2_Elemental_abundance}
We estimate the gas-phase oxygen abundance of \target\ using the strong-line diagnostic R3\,$\equiv$\,[\oiii]\,$\lambda5007$/\hb.
Other commonly used diagnostics, such as O32\,$\equiv$\,[\oiii]\,$\lambda5007$/[\oii]\,$\lambda\lambda3727,3729$, are not constraining here due to the low S/N of [\oii]\,$\lambda\lambda3727,3729$.
We apply the strong-line calibration from the JADES Survey \citep{Isobe2026},
\begin{equation}
    \log(\mathrm{R3}) = c_0 + c_1 x + c_2 x^2 + c_3 x^3
\end{equation}
where $x \equiv 12 + \log(\mathrm{O/H}) - 7.0$, $c_0 = 0.3034$, $c_1 = 0.9842$, $c_2 = -0.4987$, and $c_3 = -0.0550$.
We find that $12 + \log(\mathrm{O/H}) = 7.16 \pm 0.25$, suggesting \target\ at $\sim\!0.03\,Z_\odot$.

We further constrain the carbon-to-oxygen abundance ratio (C/O) using the
detected [\oiii]\,$\lambda5007$ and the upper limit on \ciii]\,$\lambda\lambda1907,1909$.
Following the method mentioned in \citet{Hsiao2025}, we use {\tt PyNeb} \citep{Luridiana2015}.
We estimate $\mathrm{O^{2+}/H^+}$ and $\mathrm{C^{2+}/H^+}$ based on the function {\tt getIonAbundance} from the measured fluxes of \hb\ and [\oiii]\,$\lambda5007$ and the upper limit of \ciii]\,$\lambda\lambda1907,1909$, assuming an electron temperature $T_e = 2\times10^4$\,K \citep{Nakajima2023} and electron density $n_e = 600$\,cm$^{-3}$ \citep{Li2025}, for typical $z>9$ galaxies.
We obtained $\mathrm{C^{2+}/O^{2+}} < 0.13$ and then estimated the C/O by assuming a typical ionization correction factor ICF(C/O)\,$= 1.0$, appropriate for the highly ionized conditions typical of $z \sim 9$ galaxies \citep[$\mathrm{C/O} = \mathrm{C^{2+}/O^{2+}} \times {\rm ICF}$;][]{Berg2019}.
Under this assumption, we get $\log(\mathrm{C/O}) < -0.88$.
All these estimations are summarized in Table~\ref{tab:ebg1_lines}.

\section{Discussion}
\label{sec:04_disc}

The deeper and higher-resolution spectrum of \target\ motivates updated measurements of several physical properties.
We begin by discussing the nature of \target\ through its chemical abundances (Sec. \ref{sec:EBG1_nature}).
At this metallicity, we then discuss \fesc\ by comparing the measured nebular emission lines with {\tt BEAGLE} photoionization models (Sec. \ref{sec:EBG1_escape}).
Finally, we explore the implications for cosmic reionization (Sec. \ref{sec:EBG1_reionization}).

\subsection{Gas-Phase Metallicity and C/O Ratio}
\label{sec:EBG1_nature}

\begin{figure*}
    \centering
    \includegraphics[width=0.98\linewidth]{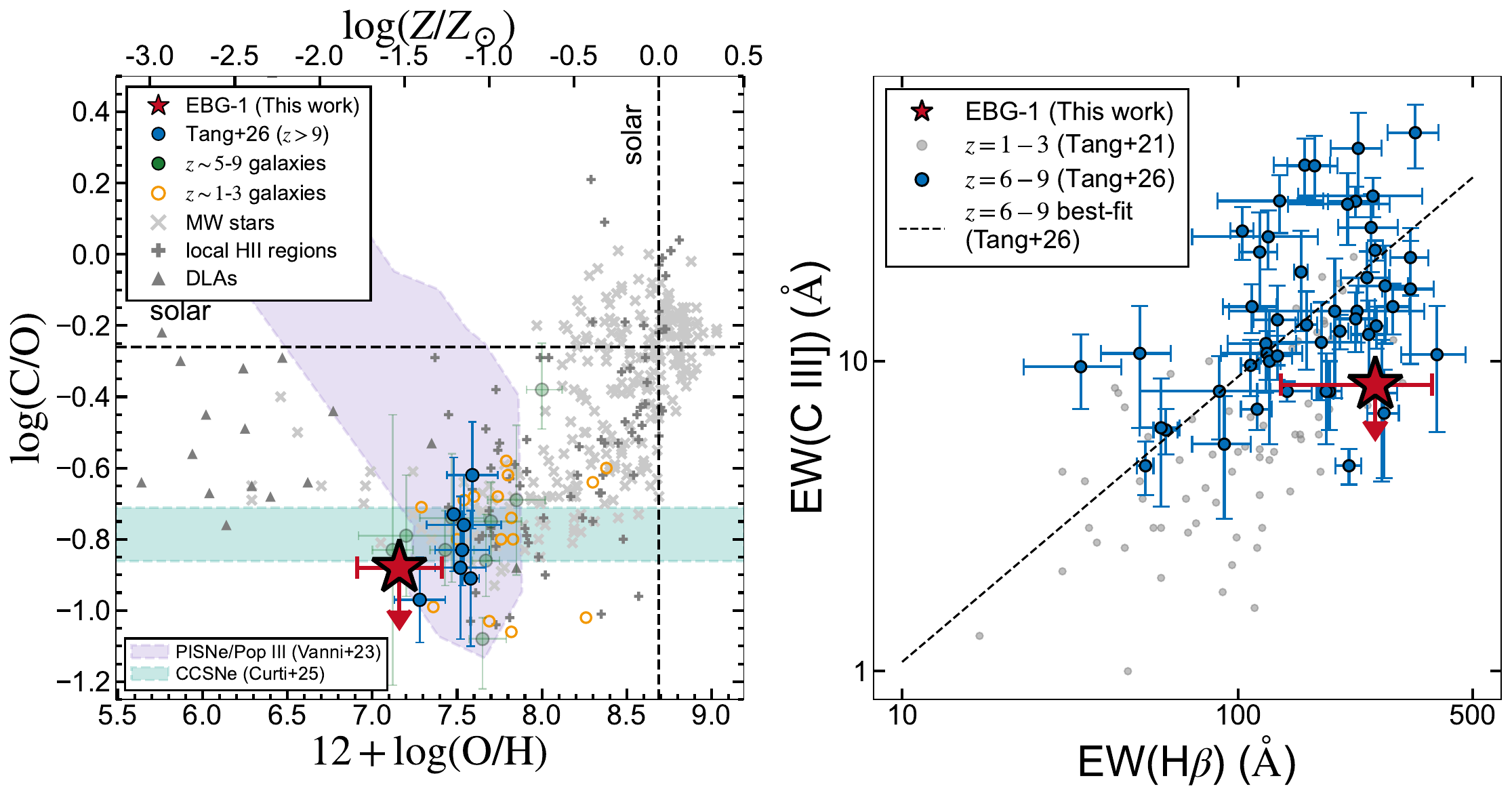}
    \caption{Chemical abundance diagnostics of \target. \textbf{Left:} The C/O-metallicity plane, with the top axis showing $\log(Z/Z_\odot)$. Shaded regions show the abundance patterns predicted for PISNe and Pop III yields \citep{Vanni2023} and for CCSNe \citep{Tominaga2007, Curti2025}. Dashed lines mark the solar abundances. The data points are adopted from \citet{Tang2026}.
    Colored symbols show galaxies at $z>9$ \citep{Tang2026}, $z\sim5-9$ \citet{ArellanoCordova2022,MarquesChaves2024,NavarroCarrera2025,Hu2024}, and $z\sim1-3$, \citep{Erb2010,Christensen2012,Bayliss2014,James2014,Stark2014,Steidel2016,Amorin2017,Berg2018,Mainali2020,Matthee2021,Iani2023,Citro2024}, while gray symbols denote Milky Way stars \citep{Gustafsson1999,Akerman2004,Bensby2006,Nissen2014}, local \hii\ regions \citep{Esteban2004,Esteban2009,Esteban2014,Garcia-Rojas2004,Garcia-Rojas2005,Garcia-Rojas2007,Lopez-Sanchez2007,Pilyugin2012,Berg2016,Berg2019,ToribioSanCipriano2016,Senchyna2017}, and DLAs~\citep{Nunez2022}. The red star marks \target\ at $12+\log(\mathrm{O/H}) = 7.16 \pm 0.25$, with the downward arrow denoting the $3\sigma$ upper limit $\log(\mathrm{C/O}) < -0.88$. \target\ following the C/O--O/H trend traced by lower-redshift galaxies to substantially lower metallicity and lies well below the PISN/Pop III region, and also the CCSN-dominated enrichment. \textbf{Right:} The EW(\ciii])--EW(H$\beta$) plane. Gray and blue circles show star-forming galaxies at $z=1-3$ \citep{Tang2021} and $z=6-9$ \citep{Tang2026}, respectively, while the dashed line shows the best-fit relation for the $z=6-9$ sample. The red star marks \target, with the downward arrow denoting the $3\sigma$ upper limit EW(\ciii])\,$< 8.4$\,\AA\ and the horizontal error bar showing the $1\sigma$ uncertainty in EW(H$\beta$). Despite its large H$\beta$ EW, \target\ falls below the \ciii]--H$\beta$ relation at fixed EW(H$\beta$), consistent with a lower carbon abundance.}
    \label{fig:EBG1_abundance}
\end{figure*}

Oxygen is produced by massive stars through core-collapse supernovae (CCSNe), but carbon is contributed by both massive stars and longer-lived intermediate-mass stars \citep{Tominaga2007}.
The C/O ratio therefore provides a sensitive diagnostic of the stellar populations and enrichment timescales that established the earliest chemical abundances.
These mechanisms produce the C/O--metallicity sequence observed in the local Universe \citep[e.g.,][]{Berg2016,Berg2019,Steidel2016}.
At the lowest metallicities, some damped Ly$\alpha$ absorbers (DLAs) show an upturn in C/O toward the solar ratio, similar to that observed in metal-poor Galactic halo stars \citep{Cooke2011,Cooke2017}. This enhancement may reflect the nucleosynthetic yields of the earliest generations of massive stars, potentially including Population III stars \citep{Vanni2023}.
Whether this excess is generic to the most metal-poor galaxies at $z>9$ is not well tested.
We explore the abundance in the left panel of Figure~\ref{fig:EBG1_abundance}, which places \target\ in the $\log(\mathrm{C/O})$--$\log(Z/Z_\odot)$ plane.

The R3-derived metallicity of \target\ is $12 + \log(\mathrm{O/H}) = 7.16 \pm 0.25$ ($\log(Z/Z_\odot) \simeq -1.5$), and the non-detection of \ion{C}{3}] implies a low carbon-to-oxygen ratio, $\log(\mathrm{C/O}) < -0.88$ ($3\sigma$).
These values place \target\ broadly along the trend of lower-redshift galaxies, in which lower-metallicity systems exhibit lower C/O.
Compared with theoretical enrichment calculations, \target\ lies at or slightly below the plateau expected from pure core-collapse supernova enrichment \citep[][]{Curti2025} and is inconsistent with the elevated C/O predicted by Pop~III enrichment models \citep{Vanni2023}.
This indicates that no distinct carbon-production mechanism is required for this extremely blue galaxy at $z > 9$; such a low C/O at low metallicity is instead the expected signature of rapid, recent star formation enriched by core-collapse supernovae.

To further understand the weak \ciii]\ emission from \target, we compare our measurement with \ciii]\ emitters in the literature \citep{Tang2021, Tang2026} in the right panel of Figure~\ref{fig:EBG1_abundance}.
At similar EW(H$\beta$), most of these galaxies are detected in \ciii], whereas \target\ falls below the best-fit EW(\ciii])--EW(H$\beta$) relation, further supporting the scenario in which \target\ is a metal-poor and carbon-poor galaxy with a high escape fraction.
We note that all carbon constraints rely on the $3\sigma$ upper limit on \ciii]$\lambda\lambda1907,1909$, and that the H$\beta$ detection remains marginal (S/N $\simeq 2.1$), which propagates into the diagnostics of \target\ in both parameter spaces.
Deeper spectroscopy reaching fainter \ciii]\ limits is required to precisely constrain C/O.

\subsection{Emission-Line Diagnostics Favor a High \fesc\ for \target}
\label{sec:EBG1_escape}

\begin{figure*}
    \centering
    \includegraphics[width=0.98\linewidth]{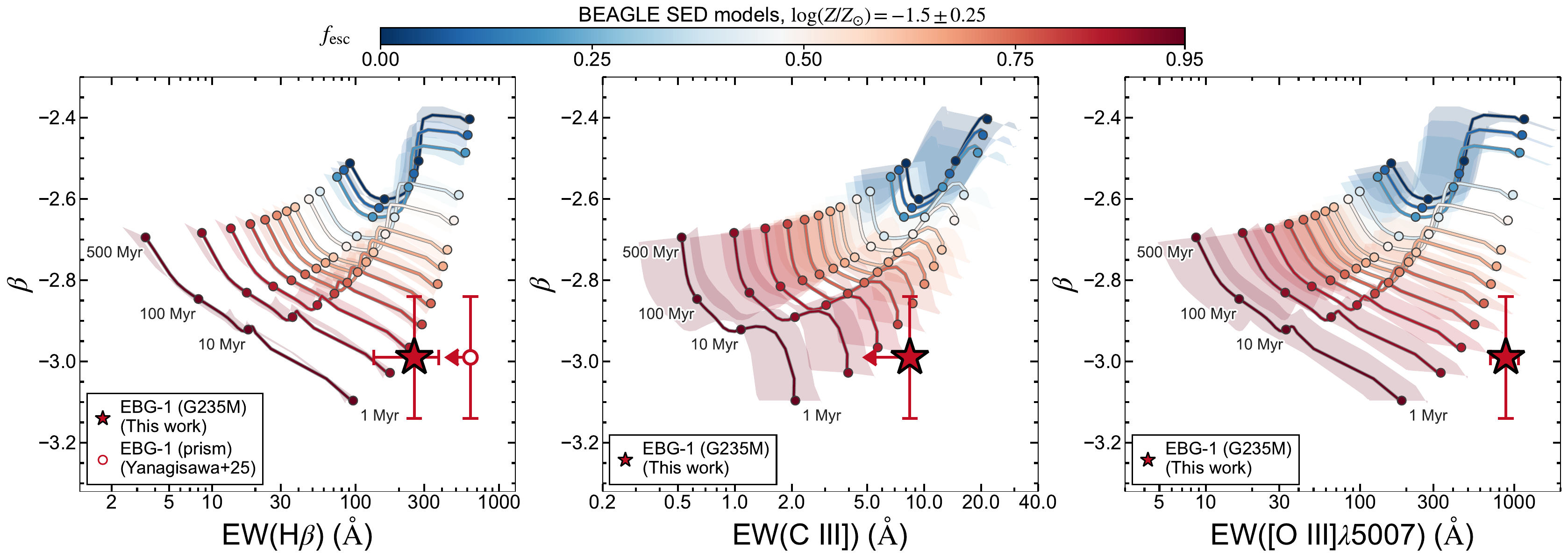}
    \caption{Escape fraction diagnostics for \target\ from nebular emission lines. In all panels, the tracks show predictions from synthetic BEAGLE spectra assuming a constant star formation history (CSFH). The models are computed over the metallicity range $\log(Z/Z_{\odot}) = -1.5 \pm 0.25$, consistent with the measured metallicity of \target (Section~\ref{sec:03_2_Elemental_abundance}). Circles mark stellar ages of 1, 10, 100, and 500\,Myr. Shaded bands indicate the model variation across the adopted metallicity range. \textbf{Left:} The $\beta$--EW(H$\beta$) plane. The open circle marks the prism-based measurement from \citet{Yanagisawa2025}, with the arrow denoting the $3\sigma$ upper limit on the prism H$\beta$ EW. The filled star marks the G235M measurement from this work. Error bars show $1\sigma$ uncertainties in all panels. \textbf{Middle:} The $\beta$--EW(\ciii]) plane.  \textbf{Right:} The $\beta$--EW([\oiii]$\lambda5007$) plane. 
    The three diagnostics consistently favor models with \fesc\,$\gtrsim 0.75$, with \fesc$\approx0.85$ providing the best overall agreement with the observations. Models with \fesc\,$\lesssim 0.5$ cannot reproduce the observed properties simultaneously.
    }
    \label{fig:EBG1_fesc}
\end{figure*}

The extremely blue UV slope of \target, $\beta = -2.99 \pm 0.15$, places it among the bluest galaxies observed at $z \gtrsim 9$ \citep{Yanagisawa2025}.
As suggested by \citet{Chisholm2022}, the extremely blue UV slope of \target\ disfavors a significant contribution from nebular continuum emission.
We therefore constrain \fesc\ of \target\ at its measured metallicity, following the framework of \citet{Topping2022} in the $\beta$--EW(H$\beta$) plane and we extending it to the $\beta$--EW(\ciii]) and $\beta$--EW(\oiii) planes.
We construct model grids with {\tt BEAGLE} \citep{Chevallard2016}.
We assume a constant star formation history, a \citet{Chabrier2003} initial mass function with an upper-mass cutoff of $300\,M_\odot$, an ionization parameter of $\log U = -2.5$ (a typical value for young stellar population following \citet{Tang2026}), and no dust attenuation.
Stellar ages span $\log(\mathrm{Age/yr}) = 6-8.7$ (1-500 Myr), and the metallicity is fixed to $\log(Z/Z_\odot) = -1.5 \pm 0.25$, matching the R3-derived value of \target.

We begin with H$\beta$, as its EW is only weakly sensitive to metallicity.
The left panel of Figure~\ref{fig:EBG1_fesc} shows the $\beta$--EW(H$\beta$) plane.
Models with \fesc\,$\lesssim 0.5$ cannot reproduce $\beta < -2.7$ at any age, and only those with \fesc\,$\gtrsim 0.75$ extend to $\beta \simeq -3.0$.
With $\beta = -2.99 \pm 0.15$ and EW(H$\beta$)\,$= 256.2 \pm 122.2$\,\AA, \target\ is broadly consistent with the \fesc\,$= 0.75$--$0.9$ tracks and favors \fesc\,$\approx 0.85$.
The prism spectrum of \citet{Yanagisawa2025} yielded only a weak upper limit on EW(H$\beta$) ($\log[\mathrm{EW(H\beta)/\text{\AA}}] < 2.8$), which left \fesc\ unconstrained.
Our tentative ($\sim 2\sigma$) detection of H$\beta$ tightens the constraints on both R3 (and hence the model metallicity) and EW(H$\beta$).

We next extend the same analysis to \ciii] and [\oiii]$\lambda5007$.
Both lines are collisionally excited and therefore depend on metallicity and ionization parameter.
We evaluate them at the metallicity measured for \target\ in Section~\ref{sec:03_2_Elemental_abundance}.
The middle panel shows the $3\sigma$ limit of EW(\ciii])\,$< 8.4$\,\AA\ at the measured $\beta$.
This non-detection is consistent with the model prediction at low metallicity and low carbon abundance.
The right panel of Figure.~\ref{fig:EBG1_fesc} shows [\oiii]$\lambda5007$, our most significant detection, with EW([\oiii]$\lambda5007$)\,$= 882.8 \pm 180.6$\,\AA\ at $5\sigma$.
All three diagnostics place \target\ in the same region of the model grid.
This region corresponds to a young, metal-poor stellar population with high \fesc.
We therefore infer \fesc\,$\gtrsim 0.75$ for \target, with \fesc\,$\approx 0.85$ most favored.
For the following analysis, we adopt the favored value of \fesc\,$= 0.85$ to estimate the intrinsic ionizing photon production efficiency and assess the contribution of \target to cosmic reionization.

\subsection{The Implications of \target\ for Cosmic Reionization}
\label{sec:EBG1_reionization}

\begin{figure*}
    \centering
    \includegraphics[width=0.98\linewidth]{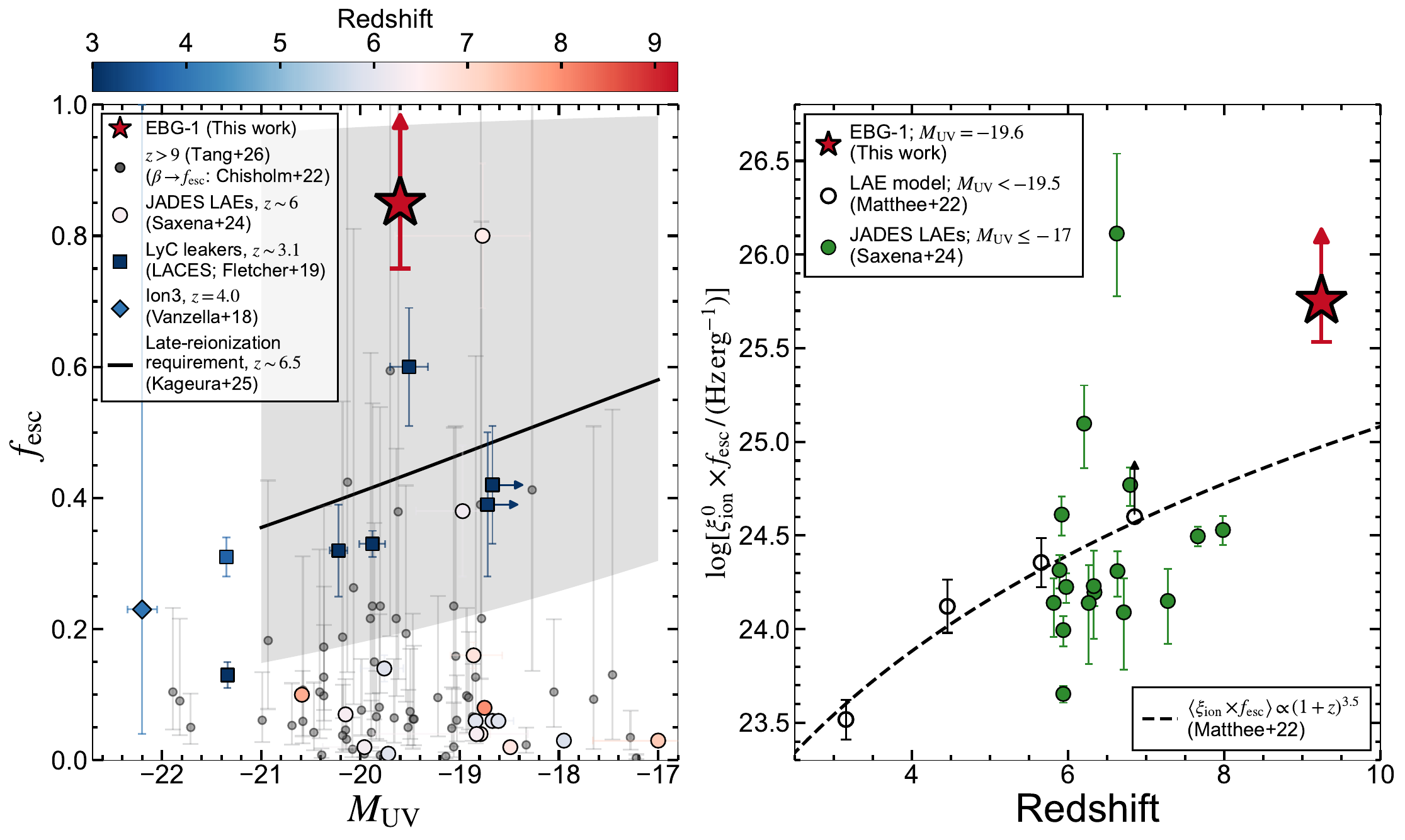}
    \caption{Impact of \target\ on cosmic reionization. \textbf{Left:} LyC escape fraction \fesc\ as a function of UV magnitude $M_{\rm UV}$. Gray circles show $z>9$ galaxies from \citet{Tang2026}, with \fesc\ inferred from their UV slopes using the \citet{Chisholm2022} relation. Light-colored circles show LAEs at $z\sim6$ from the JADES survey \citep{Saxena2024}, while blue squares and diamonds represent lower-redshift LyC leakers from \citet{Fletcher2019} and \citet{Vanzella2018}, respectively. The black solid line and gray shaded region indicate the escape fraction required by the late-reionization model of \citet{Kageura2025} at $z\sim6.5$ and its $1\sigma$ uncertainty. \target\ is marked by the red star at the favored value of \fesc$\,\approx0.85$, with the upward arrow indicating the model-supported lower bound of \fesc$\,\gtrsim0.75$. This demonstrates that relatively UV-bright galaxies can reach the exceptionally high escape fractions invoked by late-reionization scenarios. \textbf{Right:} Redshift evolution of $\xi_{\rm ion}^{0}\times$\fesc. The black circles show the LAE-driven reionization model of \citet{Matthee2022} for $M_{\rm UV}<-19.5$, and the dashed line represents their best-fit $(1+z)^{3.5}$ relation. Green circles show faint JADES LAEs at $M_{\rm UV}\leq-17$ \citep{Saxena2024}. The $\xi_{\rm ion}^{0}$ of \target\ is derived from the \hb\ emission after correcting for LyC escape. We adopt the best-favored \fesc$\,\approx0.85$ to calculate its escaping ionizing efficiency (red star), with the upward arrow indicating the broader high-escape-fraction constraint. At $z=9.244$, \target\ ($M_{\rm UV}=-19.6$) lies $\sim0.5$\,dex above the extrapolated \citet{Matthee2022} relation, indicating an exceptionally high escaping ionizing photon production efficiency.}
    \label{fig:reionization}
\end{figure*}

With $M_{\rm UV}=-19.6$, \target\ is $\sim30$ times more UV-luminous than the faint ($M_{\rm UV}>-16$) galaxies predicted to dominate the ionizing photon budget in some reionization models \citep{Finkelstein2019}.
However, the relative contributions of faint and luminous galaxies to cosmic reionization remain debated, largely because their ionizing escape fractions are poorly constrained \citep{Naidu2020,Harikane2020}.
The left panel of Figure~\ref{fig:reionization} compares the inferred \fesc\ of \target\ with $z>9$ galaxies from \citet{Tang2026}, whose escape fractions are estimated from their UV slopes using the \citet{Chisholm2022} relation.
Also, we compared faint LAEs from the JADES survey \citep{Saxena2024}.
Compared with these samples, \target\ exhibits an exceptionally high escape fraction of \fesc$\,\approx0.85$ at its UV luminosity.
Such high escape fractions are particularly relevant to late reionization scenarios, which require efficient ionizing sources to complete reionization over a short cosmic time.
For example, the model of \citet{Kageura2025} requires escape fractions approaching \fesc$\,\sim0.5$ at $z\sim6.5$ if massive halos dominate the ionizing photon budget.
Although this model applies to a lower redshift, \target\ provides evidence that relatively UV-bright galaxies can achieve the high escape fractions invoked by these scenarios.
We therefore estimate the escaping ionizing photon production efficiency, \fesc$\,\times\,\xi_{\rm ion}^{0}$, of \target\ to assess its potential contribution to cosmic reionization.

Following \citet{Matthee2023}, we estimate the ionizing photon production efficiency as
\begin{equation}
    \xi_{\rm ion} = \frac{L_{{\rm H}\beta}}{c_{{\rm H}\beta}\,L_{\rm UV}},
\end{equation}
where $L_{{\rm H}\beta}$ is the \hb\ luminosity, $L_{\rm UV}$ is the UV luminosity derived from the measured $M_{\rm UV}$, and $c_{{\rm H}\beta} = 4.86 \times 10^{-13}$\,erg is the line-emission coefficient for case B recombination.
We assume negligible dust attenuation, consistent with the blue UV slope of \target\ \citep{Yanagisawa2025}.
Adopting the magnification-corrected \hb\ and UV luminosities, we obtain $\log(\xi_{\rm ion}/{\rm Hz\,erg^{-1}}) = 25.00 \pm 0.21$, assuming zero LyC escape.
Correcting for escaping ionizing photons, we define the intrinsic efficiency as $\xi_{\rm ion}^{0} \equiv \xi_{\rm ion}/(1-$\fesc$)$.
Adopting \fesc$\,\approx0.85$, we infer $\log(\xi_{\rm ion}^{0}/{\rm Hz\,erg^{-1}}) \approx 25.82$ and $\log(\xi_{\rm ion}^{0}\times$\fesc$/{\rm Hz\,erg^{-1}}) \approx 25.75$.
The inferred intrinsic efficiency exceeds the canonical value of $\log(\xi_{\rm ion}/{\rm Hz\,erg^{-1}}) = 25.2$ \citep{Robertson2013} by approximately $\sim0.5$\,dex.
In the right panel of Figure~\ref{fig:reionization}, we compare the escaping ionizing efficiency of \target\ with the LAE-driven reionization model of \citet{Matthee2022}, computed for $M_{\rm UV} < -19.5$, and with faint LAEs from the JADES survey \citep{Saxena2024}.
At $z=9.244$, \target\ lies $\approx0.5$\,dex above the extrapolated population-averaged relation of \citet{Matthee2022}.
This comparison suggests that \target\ may represent an exceptionally efficient ionizing source of the type considered in LAE-driven reionization scenarios.
We caution that \target\ is not a confirmed LAE, and its Ly$\alpha$ properties remain unconstrained.

Although \target\ exhibits an exceptionally high escaping ionizing efficiency, its contribution to cosmic reionization depends on the abundance of galaxies with similar properties.
Following the method of \citet{MarquesChaves2026}, we estimate the fraction ($f_N$) of the total UV luminosity density contributed by galaxies with \target-like ionizing properties required to match the escaping ionizing emissivity of a typical galaxy population.
We adopt $f_{\rm esc,typical}=15\%$ and $\log(\xi_{\rm ion,typical}^{0}/\mathrm{Hz\,erg^{-1}})=25.25$ for the typical population.
The fraction $f_N$ is then:
\begin{equation}
    f_N
    = \frac{f_{\rm esc,typical}\,\xi_{\rm ion,typical}^{0}}
    {f_{\rm esc}\,\xi_{\rm ion}^{0}}
    \approx \frac{0.15\times10^{25.25}}
    {0.85\times10^{25.82}}
    \approx 0.05.
\end{equation}
Under these assumptions, galaxies with \target-like ionizing properties contributing only $\sim5\%$ of the total UV luminosity density could match the escaping ionizing emissivity of a typical galaxy population.

\section{Summary}
\label{sec:05_sum}

We present deep (14\,hr) JWST/NIRSpec G235M ($R \sim 1000$) spectroscopy of \target, a gravitationally lensed galaxy at $z = 9.244$ that lies in the bluest regime among galaxies observed at $z > 9$.
The medium-resolution spectrum yields significantly improved constraints on the \ciii], H$\beta$, and [\oiii] emission lines relative to the previous prism spectroscopy \citep{Yanagisawa2025}.
Based on these measurements, we characterize the nature of \target\ and its escape fraction in detail.
Our main results are summarized as follows.
\begin{enumerate}
\item
The deep rest-frame UV-to-optical spectrum of \target\ is characterized by weak emission lines.
We place a $3\sigma$ upper limit of EW(\ciii]) $< 8.4$\,\AA, together with a tentative ($\sim 2\sigma$) detection of H$\beta$ (EW $= 256.2 \pm 122.2$\,\AA) and a $5\sigma$ detection of [\oiii]\,$\lambda 5007$ (EW $= 882.8 \pm 180.6$\,\AA).
From the \ciii]/[\oiii] line ratio and the R3 index ([\oiii]\,$\lambda 5007$/H$\beta$), we conclude that \target\ is a metal-poor galaxy with $12 + \log({\rm O/H}) = 7.16 \pm 0.25$ ($\simeq 3\%\,Z_{\odot}$) and a carbon-to-oxygen abundance of $\log({\rm C/O}) < -0.88$ ($3\sigma$), following the C/O--metallicity trend of lower-redshift galaxies.
\item
\target\ shows spectroscopic signatures of a high escape fraction, a quantity that has been poorly constrained for extremely blue $z > 9$ galaxies to date.
Combining the observed EW(H$\beta$) and EW(\ciii]) with the UV slope $\beta = -2.99 \pm 0.15$, and comparing with BEAGLE models computed at the measured metallicity, \target\ lies in the model regime of \fesc$\,\gtrsim 0.75$ (\fesc$\,\approx 0.85$) in both the $\beta$--EW(H$\beta$) and $\beta$--EW(\ciii]) planes.
\target\ also falls below the \ciii]--H$\beta$ relation of $z = 6$--$9$ galaxies at fixed EW(H$\beta$), consistent with substantial leakage of ionizing photons.
\item
To assess the implications of \target\ for late reionization, we compare its inferred LyC escape fraction with the model of \citet{Kageura2025}.
We further compare its escaping ionizing photon production efficiency (\fesc$\,\times\,\xi_{\rm ion}^{0}$) with the LAE-driven model of \citet{Matthee2022} and faint JADES LAEs \citep{Saxena2024}.
The BEAGLE diagnostics favor \fesc\,$\gtrsim 0.75$, with \fesc\,$\approx 0.85$ preferred. Adopting \fesc\,$\approx 0.85$, we find that \target\ reaches the high escape fractions invoked by late-reionization scenarios. Its escaping ionizing efficiency lies $\sim 0.5$ dex above the extrapolated population-averaged relation of \citet{Matthee2022}.
Our ionizing-budget estimate suggests that galaxies like \target\ would be only $\sim5\%$ to match the escaping ionizing emissivity predicted for the typical population.
\end{enumerate}

Recent JWST observations have begun to connect galaxy overdensities with enhanced Ly$\alpha$ transmission and the growth of ionized regions at $z\simeq7-8.5$ \citep{Chen2026}.
This connection remains less constrained at $z\simeq9$ despite recent studies of candidate ionized regions \citep{Whitler2024,Whitler2026}.
\target\ offers a complementary opportunity to study this connection through a galaxy with an extremely blue UV continuum and a high inferred LyC escape fraction.
Future observations of its galaxy environment would test whether \target\ resides in an overdense region or a more typical environment.
A census of neighboring galaxies would help assess whether the local ionizing budget could be dominated by \target\ or by its surroundings.
Combining this census with constraints on Ly$\alpha$ transmission and the duration of ionizing activity would help assess their relative contributions to the growth of a local ionized region.

\begin{acknowledgments}

The authors thank St\'{e}phane Charlot and Jacopo Chevallard for providing access to the {\tt BEAGLE} tool. Y. Wu thanks J. Xavier Prochaska and Takumi Tanaka for helpful discussion.

MO acknowledges the supports from the World Premier International
Research Center Initiative (WPI Initiative), MEXT, Japan, the joint research program of the Institute
for Cosmic Ray Research (ICRR), the University of Tokyo, and KAKENHI (21H04467, 25H00674) 
through JSPS.
YL acknowledges the support from the JSPS KAKENHI grant No. 24K17084 and 25H00663.

This work is based on observations made with the NASA/ESA/CSA James Webb Space Telescope. The data were obtained from the Mikulski Archive for Space Telescopes at the Space Telescope Science Institute, which is operated by the Association of Universities for Research in Astronomy, Inc., under NASA contract NAS 5-03127 for JWST. These observations are associated with program \#1433, and 7390.
The specific observations analyzed can be accessed via \dataset[doi:10.17909/aqj4-f374]{https://doi.org/10.17909/aqj4-f374}.
Support for US investigators in program \#7390 was provided by NASA through a grant from the Space Telescope Science Institute, which is operated by the Association of Universities for Research in Astronomy, Inc., under NASA contract NAS5-03127.
Kavli IPMU was established by World Premier International Research Cen-ter Initiative (WPI), MEXT, Japan. This work was performed at the Center for Data-Driven Discovery, Kavli IPMU (WPI), and made use of computing resources at Kavli IPMU.
The authors acknowledge the use of ChatGPT (OpenAI, GPT-5.6), and Claude (Opus 5 and Fable 5) to assist with language editing, code development, and figure generation. All AI-assisted outputs were carefully reviewed and validated by the authors. The authors take full responsibility for all analyses, interpretations, and conclusions presented in this work.

\end{acknowledgments}





%
\facilities{JWST}

\software{{\sc Astropy}:\footnote{\url{http://www.astropy.org}} a community-developed core Python package and an ecosystem of tools and resources for astronomy \citep{astropy2013, astropy2018, astropy2022}; {\tt BEAGLE} \citep{Chevallard2016}; {\sc Emcee} \citep{Foreman-Mackey2013}; {\sc Jupyter} \citep{Kluyver2016}; {\sc Matplotlib} \citep{Hunter2007}; {\sc Msaexp} \citep{Brammer2023zndo...8319596B}; {\sc Numpy} \citep{harris2020array}; {\sc Scipy} \citep{2020SciPy-NMeth}.
}

\appendix

\section{Consistency Between NIRCam Photometry and NIRSpec Spectrophotometry}
\label{Appendix_phot}

\begin{figure*}
    \centering
    \includegraphics[width=0.98\linewidth]{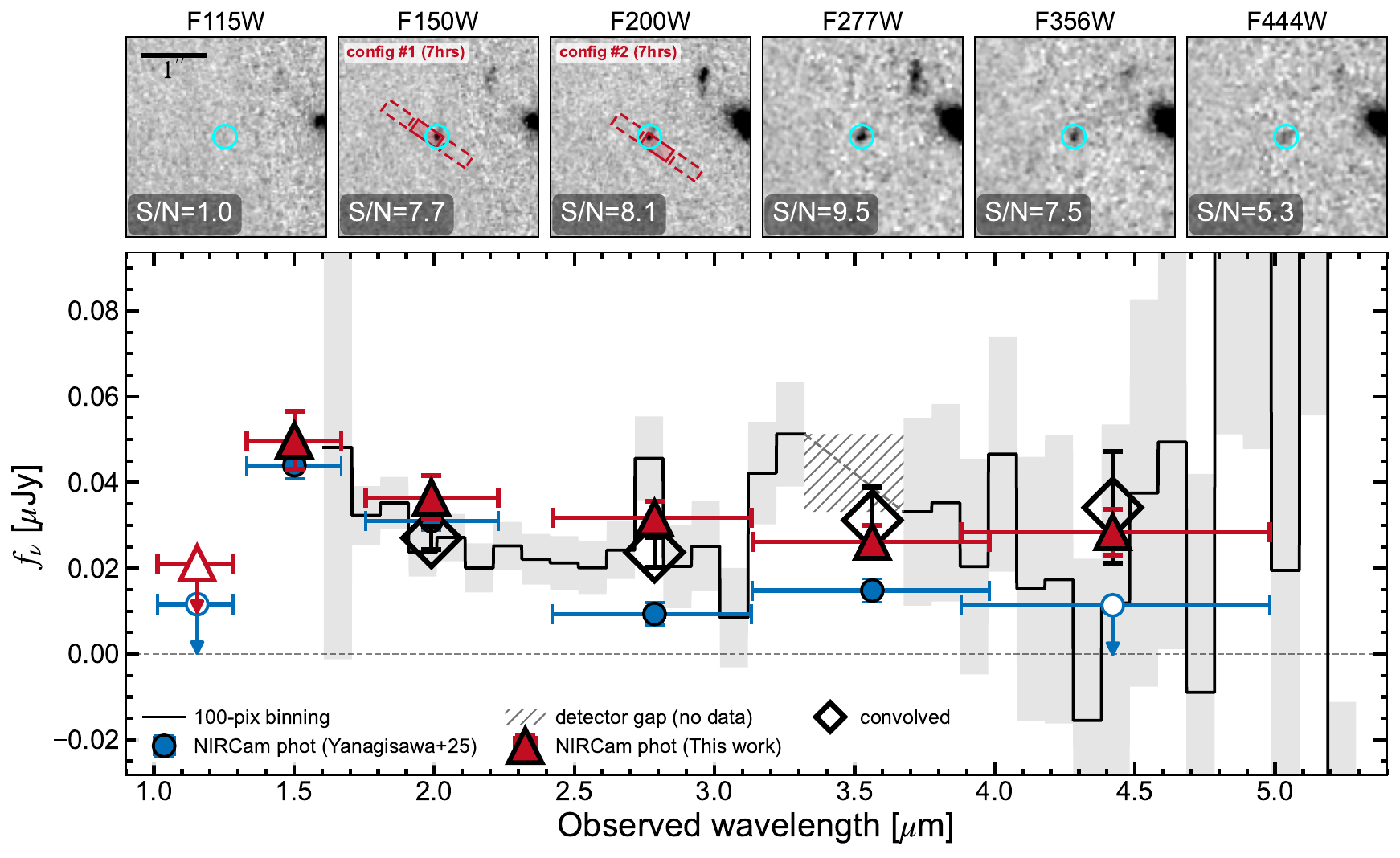}
    \caption{Comparison between the NIRSpec spectrum of \target\ and the NIRCam photometry. \textbf{Top:} NIRCam cutouts in F115W, F150W, F200W, F277W, F356W, and F444W. Cyan circles mark the $0\farcs35$-diameter circular aperture used for photometry, and the S/N of each measurement is annotated. \textbf{Bottom:} The NIRSpec G235M spectrum binned by 100 pixels (black histogram). Synthetic photometry is derived by convolving the spectrum with each NIRCam filter transmission curve, and is shown as open black diamonds with 1$\sigma$ error bars. The updated NIRCam photometry (red triangles) is compared with the values reported by \citet{Yanagisawa2025} (blue circles), where horizontal bars indicate the filter bandwidths. Open symbols with downward arrows denote 3$\sigma$ upper limits. Revisiting the photometry, we find that \citet{Yanagisawa2025} underestimated the fluxes between 3 and 5\,\micron, and the synthetic photometry supports our revised values.}
    \label{fig:spec_phot}
\end{figure*}

To check our flux calibration, we compare the fluxes derived from the NIRSpec spectrum with our NIRCam photometry and with that reported in \citet{Yanagisawa2025}. We measure the NIRCam photometry of \target\ in six bands following the procedures described in Section~\ref{sec:02_obs}, and list the results in Table~\ref{tab:phot}. \target\ is detected at S/N\,$=$\,7.7, 8.1, 9.5, 7.5, and 5.3 in F150W, F200W, F277W, F356W, and F444W, and remains undetected in F115W, as expected for a Lyman break at $z=9.244$. We then convolve the flux-calibrated 1D spectrum with each NIRCam filter transmission curve to construct synthetic photometry. Figure~\ref{fig:spec_phot} compares the two sets of measurements.
Our imaging and spectroscopy are consistent. The synthetic fluxes agree with our NIRCam photometry within $\approx 1.7\sigma$ in F200W, F277W, F356W, and F444W.

The photometry reported by \citet{Yanagisawa2025} is not uniformly consistent with either dataset. Their F150W and F200W fluxes agree with our updated values to within $1\sigma$ ($0.13$ and $0.18$\,mag, respectively), indicating that the rest-frame UV continuum and UV slope is robustly recovered in both analyses.
However, the F277W, F356W, and F444W fluxes are instead fainter than our updated measurements by $1.33$, $0.62$, and $1.93$\,mag, corresponding to $4.8\sigma$, $2.5\sigma$, and $3.6\sigma$.
We adopt our revised photometry throughout this Letter.

\section{SED Fitting and the Mass-Metallicity Relation}
\begin{figure*}
    \centering
    \includegraphics[width=0.98\linewidth]{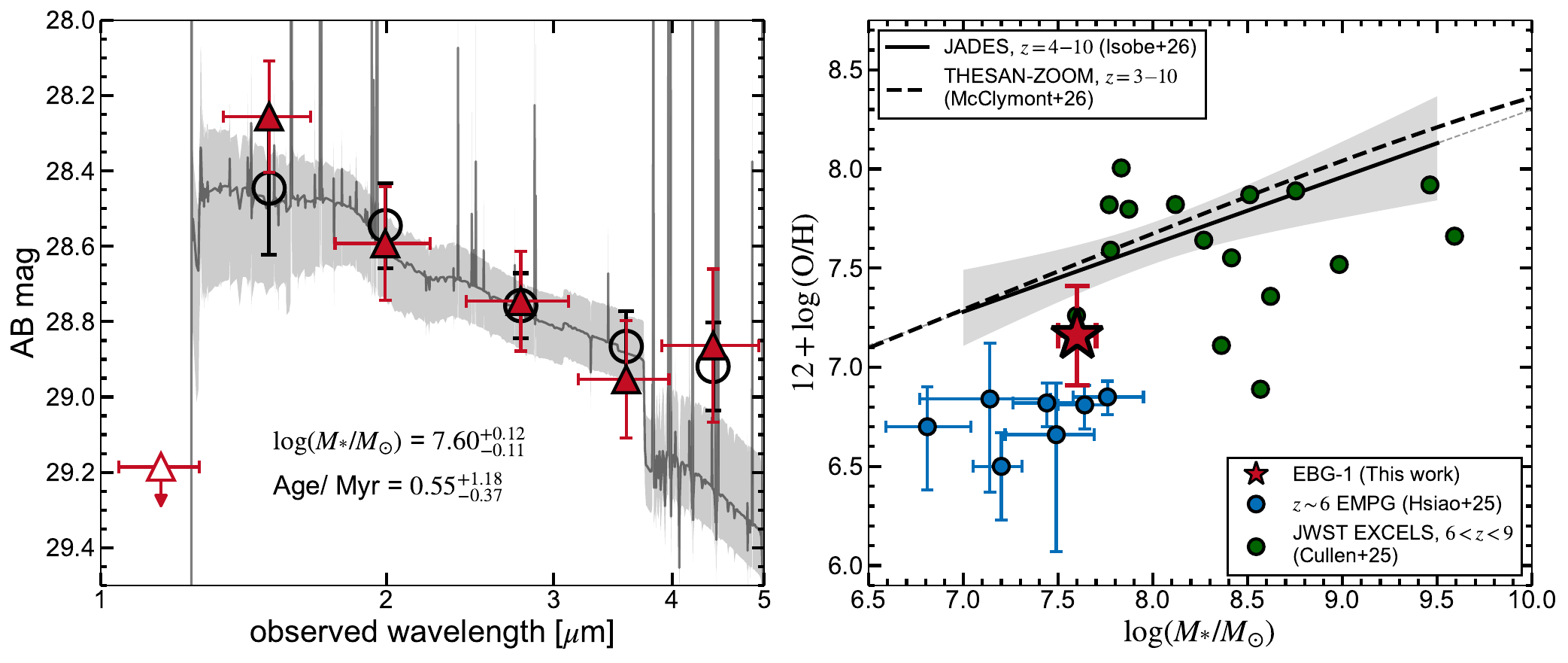}
    \caption{SED fitting results and mass--metallicity relation (MZR) of \target. {\bf Left:} {\tt BEAGLE} SED fit to the NIRCam photometry of \target\ with \fesc\ as a free parameter. The gray curve and shaded region show the best-fit model spectrum and its 68\% confidence interval, and open circles show the corresponding synthetic photometry. Red triangles show the observed photometry, and the open triangle marks the 3$\sigma$ upper limit in F115W. The fit yields $\log(M_*/M_\odot)=7.60^{+0.12}_{-0.11}$ and an age of $0.55^{+1.18}_{-0.37}$\,Myr. {\bf Right:} MZR of \target\ compared with the literature. The solid line shows the best-fit relation for JADES galaxies at $z=4-10$ \citep{Isobe2026}, and the dashed line shows the prediction of the {\sc thesan-zoom} simulations at $z=3-10$ \citep{McClymont2026}. The red star shows \target, blue circles show extremely metal-poor galaxies at $z\sim6$ \citep{Hsiao2025b}, and green circles show JWST/EXCELS galaxies at $6<z<9$ \citep{Cullen2025}. \target\ lies below the best-fit MZR at its stellar mass and falls close to the extremely metal-poor population at $z\sim6$.}
    \label{fig:SED_MZR}
\end{figure*}

We fit the revised NIRCam photometry of \target\ with \texttt{BEAGLE} \citep{Chevallard2016} to constrain its physical properties.
We fix the redshift at $z=9.244$ and adopt a \citet{Chabrier2003} initial mass function with an upper mass cutoff of $300\,M_\odot$.
We assume a constant star formation history (CSFH) and leave \fesc\ as a free parameter.
The fit yields a stellar mass of $\log(M_*/M_\odot)=7.60^{+0.12}_{-0.11}$ and a stellar age of $0.55^{+1.18}_{-0.37}$\,Myr.
These values support the picture that \target\ is a young galaxy with a high \fesc. 
The best-fit SED results are shown in the left panel of Figure~\ref{fig:SED_MZR}.

The stellar mass allows us to place \target\ on the mass--metallicity relation (MZR) established at $z=4-10$.
We combine it with the gas-phase metallicity measured in Section~\ref{sec:03_2_Elemental_abundance}.
As shown in the right panel of Figure~\ref{fig:SED_MZR}, \target\ lies below the MZR at its stellar mass.

\begin{deluxetable}{cc}
    \centering
    \tablecaption{Photometry and Properties of EBG-1}
    \tablehead{
    \colhead{Quantity} & \colhead{Value}
    }
    \startdata
    F115W & $<7.7 \, \mathrm{nJy}$ \\
    F150W & $18.1\pm2.5 \, \mathrm{nJy}$ \\
    F200W & $13.3\pm1.8 \, \mathrm{nJy}$ \\
    F277W & $11.5\pm1.4 \, \mathrm{nJy}$ \\
    F356W & $9.5\pm1.4 \, \mathrm{nJy}$ \\
    F444W & $10.3\pm1.9 \, \mathrm{nJy}$ \\
    $\log M_*/\mathrm{M_\odot}$ & $7.60^{+0.12}_{-0.11}$ \\
    Age & $0.55^{+1.18}_{-0.37}$ $\mathrm{Myr}$ \\
    \enddata
    \tablecomments{The fluxes and the stellar mass reported in this table are corrected for the lensing magnification factor $\mu=2.75$. The F115W flux is a $3\sigma$ upper limit.}
    \label{tab:phot}
\end{deluxetable}

\bibliography{sample701}{}
\bibliographystyle{aasjournalv7}



\end{CJK*}
\end{document}